\pdfoutput=1

\documentclass[11pt]{article}

\usepackage[preprint]{acl}

\usepackage{times}
\usepackage{latexsym}

\usepackage[T1]{fontenc}

\usepackage[utf8]{inputenc}

\usepackage{microtype}

\usepackage{inconsolata}

\usepackage{graphicx}

\usepackage{hyperref}

\usepackage{array}
\usepackage{amstext}
\usepackage{amsmath}
\usepackage{amssymb}
\usepackage{wrapfig}
\usepackage{enumerate}
\usepackage{paralist}
\usepackage{xspace}
\usepackage{color}
\usepackage{times}
\usepackage[colorinlistoftodos,prependcaption,textsize=normal]{todonotes}
\usepackage{pdfpages}
\usepackage{framed,enumitem}
\usepackage{multirow} 
\usepackage{tabularx}
\usepackage{booktabs}
\usepackage{listings}
\usepackage[nottoc,numbib]{tocbibind}
\usepackage{enumitem}
\usepackage{multirow}
\usepackage{subcaption}
\usepackage{svg}
\usepackage{framed}
\usepackage{tabularx}
\usepackage{multirow}
\usepackage{caption}
\usepackage{nameref}

\usepackage{makecell}

\usepackage{cancel}
\usepackage[normalem]{ulem}

\newcommand{\SECintroduction}{Introduction}
\newcommand{\SECrw}{Background and Related Work}
\newcommand{\SECpedagogicalshortcomings}{Pedagogical shortcomings of LLMs}
\newcommand{\SECllmsfortutoring}{LLMs for Tutoring}
\newcommand{\SECpf}{Productive Failure (PF)}
\newcommand{\SECmethodology}{Methodology}
\newcommand{\SECformalism}{Formalism}
\newcommand{\SECdialogtutoringtask}{Dialog Tutoring Task}
\newcommand{\SECtutoringstrategymodeling}{Tutoring Strategy Modeling}
\newcommand{\SECstudentstatespace}{Student State Tracing}
\newcommand{\SECalgorithm}{Algorithm}
\newcommand{\SECexperiments}{Experiments}
\newcommand{\SECrqs}{Research Questions}
\newcommand{\SECdataset}{Dataset}
\newcommand{\SECbaselines}{Baselines}
\newcommand{\SECevaluationmetrics}{Evaluation Metrics}
\newcommand{\SECsyntheticstudy}{Preliminary Study with Simulated Students}
\newcommand{\SECfieldtest}{Field Test with Students}
\newcommand{\SECfeedbackquestionnaire}{Post-study Questionnaire}
\newcommand{\SECresultsdiscussion}{Results and Discussions}
\newcommand{\SECresultssyntheticstudy}{Preliminary Study with Simulated Students}
\newcommand{\SECresultsfieldtest}{Field Test with Students: Measuring end-to-end Effectiveness}
\newcommand{\SECresultsfidelity}{\textbf{RQ1.} Tutoring Strategy Fidelity}
\newcommand{\SECresultsfeedback}{\textbf{RQ2.} Student Feedback Questionnaire}
\newcommand{\SECdiscussions}{Questionnaire Feedback}
\newcommand{\SECconclusionslimitations}{Conclusion}
\newcommand{\SECappendix}{Appendix Learning Sciences and Implementation Details}
\newcommand{\SECappendiximplementation}{Prompts and Hyperparameters}
\newcommand{\SECappendixassessor}{Student State Tracing}

\newcommand{\SECappendixpromptgenerator}{Intent Dependent Steering}
\newcommand{\SECappendixtaxonomy}{Intents Taxonomy}
\newcommand{\SECtestproblems}{Test Problems}
\newcommand{\dqone}{\textbf{DQ1}}
\newcommand{\SECappendixintentselectorprompt}{V4 LLM Intent Selection Prompt}
\newcommand{\SECappendixstudentstatetracingtest}{Student State Tracing Evaluation}

\newcommand{\SECaccuracystudentstatetracing}{Accuracy of Student State Tracing}

\newcommand{\SECappendixdataset}{Dataset}

\newcommand\name[0]{StratL}
\newcommand\nproblems[0]{15}

\newcommand{\applyEdits}{} 
\newcommand{\hlred}[1]{\ifdefined\applyEdits\relax\else\textcolor{red}{#1}\fi}

\title{Towards the Pedagogical Steering of Large Language Models for Tutoring:
A Case Study with Modeling Productive Failure}

\author{
 \textbf{Romain Puech\textsuperscript{1,2}},
 \textbf{Jakub Macina\textsuperscript{1,3}},
 \textbf{Julia Chatain\textsuperscript{4}},
 \textbf{Mrinmaya Sachan\textsuperscript{1}},
 \textbf{Manu Kapur\textsuperscript{4}}
\\
\\
 \textsuperscript{1}ETH Zurich,
 \textsuperscript{2}\'Ecole polytechnique,
 \textsuperscript{3}ETH AI Center,
 \\
 \textsuperscript{4}Professorship for Learning Sciences and Higher Education, ETH Zurich,
\\
 \small{
   \href{mailto:romain.puech@alumni.polytechnique.org}{romain.puech@alumni.polytechnique.org}
 }
}

\begin{document}
\maketitle

\begin{abstract}
One-to-one tutoring is one of the most efficient methods of teaching. %
With the growing popularity of Large Language Models (LLMs), there have been efforts to %
create LLM-based conversational tutors which can expand the benefits of one-to-one tutoring to everyone. However, current LLMs are trained primarily to be helpful assistants and lack crucial pedagogical skills. For example, they often quickly reveal the solution to the student and fail to plan for a richer multi-turn pedagogical interaction.
To use LLMs in pedagogical settings, they need to be steered to use effective teaching strategies: a problem we introduce as \textit{Pedagogical Steering}.
We %
develop \name, an algorithm to optimize LLM prompts and
steer it to follow a predefined multi-turn tutoring plan represented as a transition graph. %
As a case study, we create a prototype tutor for high school math following \textit{Productive Failure} (PF), an advanced and effective learning design.
To validate our approach in a real-world setting, we run a field study with 17 high school students in Singapore %
and show that \name\ succeeds in steering the LLM to follow the PF tutoring strategy. %
Finally, we highlight challenges in Pedagogical Steering of LLMs and offer opportunities for further improvements %
by publishing a dataset of PF problems and our code\footnote{Dataset \href{https://github.com/eth-lre/Productive-Failure-Problems/tree/main}{hyperlink}; Code 
 \href{https://github.com/RomainPuech/StratL-Pedagogical-Steering-of-LLMs-for-Tutoring}{hyperlink}}.

\end{abstract}

\begin{figure*}[h!]
    \centering
    \includegraphics[width=.9\textwidth]{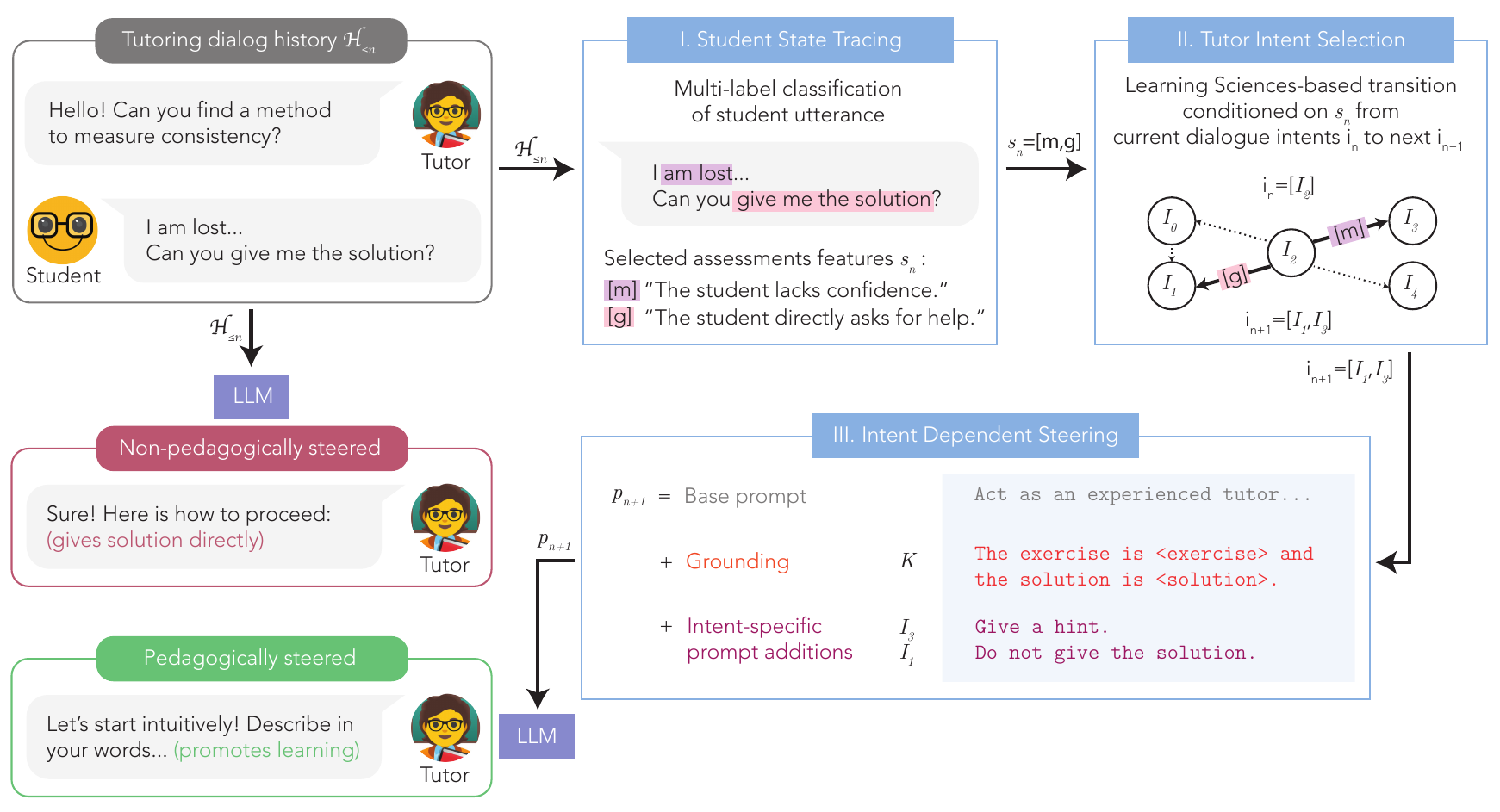}
    \caption{Schematic representation of \name\ (in blue), an algorithm to control an LLM's tutoring strategy.
    }
    \label{fig:main}
    \vspace{-1.2em}
\end{figure*}

\section{\SECintroduction{}}\label{introduction}
One-to-one tutoring is an effective teaching method~\cite{Bloom1984The2S}.
Because of the efficiency of tutoring and its prohibitive cost, the challenge of finding scalable and affordable one-to-one tutoring solutions is considered as one of the greatest problems in education: the so-called \textit{Two-Sigma Problem}~\cite{Bloom1984The2S,meta_analysis_tutoring}.
Following the rising popularity of Large Language Models (LLMs), there has been a growing interest in using them to make tutoring affordable by building conversational tutoring systems~\cite{learnLM, khanmigo}.
However, previous research has highlighted the lack of pedagogical properties of LLMs~\cite{opportunities,AI_Teacher_Test}.
Most LLMs are instruction-tuned to optimize for a broad set of human preferences~\cite{RLHF,instruction_tuning}, leading to assistant-like responses. However, a tutor's goal is to maximize learning, not user satisfaction. 

These two goals can be contradictory. For example, if a student asks for help, providing them directly with the solution would likely maximize their satisfaction and be the response preferred by LLMs~\cite{mathdial}. Yet, it is more efficient from a learning standpoint to promote students' active engagement and let them cognitively engage with the problem, for example, by providing an indirect hint or asking a guiding question~\cite{active_learning,efficient_techniques}. Moreover, existing LLM tutors are optimized and mostly evaluated for single-turn requests, while a tutoring dialog should be envisioned as a multi-turn teacher-student interaction where the teacher uses various pedagogical cues~\cite{efficient_techniques}. As LLMs are not natively made for tutoring, we introduce the shift from their user-serving goal to a pedagogically suitable one as the \textit{Pedagogical Steering} problem.

We present and evaluate \name, an algorithm to model a multi-turn \textit{tutoring strategy} with LLMs. We define a multi-turn strategy as a succession of single-turn pedagogical goals called tutoring \textit{intents}. \name\ uses a transition graph to dynamically redefine the tutoring intents after every student utterance and prompts the LLM to follow these intents for the next turn. We partner with learning scientists to integrate a strategy based on an advanced learning design known as Productive Failure (PF)~\cite{productive_failure,kapur2024productive}, which is more effective for human learning than other classical teaching methods~\cite{meta_analysis_pf}.
PF can be seen as the most extreme version of not directly telling the answer to the students. In PF, students are encouraged to explore an ill-structured problem, for example by using Socratic questioning, and are confronted with various suboptimal solutions, even if it leads to failure. While PF works across learning subjects, this work focuses on high school math, as it has been most explored in PF research~\cite{meta_analysis_pf}.

We evaluate \name\ and uncover practical limitations and insights on the Pedagogical Steering problem through a preliminary study with simulated students and a field study with 17 high school students in Singapore. Our results show that \name\ effectively steers the LLM to adhere to the PF tutoring strategy and does so with no negative spillover effects on other desirable properties of the LLM.

We summarize our contributions as follows:
\begin{itemize}
\item We define the problem of \textit{Pedagogical Steering} in the context of LLM multi-turn tutoring. \hlred{We explain why instruction-tuned LLMs are misaligned for tutoring from a Learning Sciences standpoint, and review the current state of research from this perspective. 
\item} We introduce the \name\ algorithm to model the concept of \textit{tutoring strategy} and to prompt an LLM towards a given strategy using a sequence of single-turn intents. 
\item We openly release the first dataset of PF tutoring exercises. These open-ended exercises have multiple suboptimal solutions and require multiple interactions to find the best one. 
\item We run a field test in a Singaporean high school to evaluate the potential real-world impact of \name. We measure the extent to which \name\ and its ablations control the pedagogy \hlred{followed by the LLM} through novel Learning Sciences--inspired metrics and investigate the limitations.
\end{itemize}

\section{\SECrw{}}\label{background}

\subsection{\SECpedagogicalshortcomings{}\label{sec:pedagogicalshortcomings}}
Recent advances enable state-of-the-art LLM-based systems to reliably perform middle-- to high school--level mathematics, such as using prompting~\cite{CoT,zero_shot}, multiple sampling strategies with consistency sampling~\cite{selfconsistency} or refinement~\cite{madaan2024self,ART_refinement,DiVeRSe}, fine-tuning~\cite{STaR_bootstrapping,finetuning_scaling} or use of external tools like symbolic solvers or code interpreters~\cite{symbolic,ALM}. 
\hlred{These reasoning capabilities could give LLMs the ability to serve as a basis for lesson-agnostic tutors.}

Notwithstanding the promising capabilities of LLMs, several studies exhibited their poor pedagogical properties~\cite{AI_Teacher_Test}. Among other flaws, most models often reveal the solution of the exercise to the students, even if prompted not to do so~\cite{mathdial}. \hlred{Their inefficiency as learning tools has also been measured more directly: when comparing students who were asked to solve problems using GPT-generated hints or human-generated ones, the human-generated hints group achieved substantially and statistically significantly higher learning gains~\cite{learning_gains_hint_gpt}.

}These pedagogical limitations stem from a core problem: LLMs are not optimized for tutoring. \hlred{At a fundamental level, autoregressive LLMs are transformer-based models trained to take a sequence of tokens as input and predict the most likely following one~\cite{what_is_LLM}.} After their initial pre-training, an \textit{alignment} step is necessary to turn base language models into useful assistants capable of following human instructions, like ChatGPT. The alignment is often performed by fine-tuning LLMs on human preferences~\cite{instruction_tuning}, mainly using a method known as Reinforcement Learning with Human Feedback (RLHF)~\cite{RLHF}. During this process, LLMs learn to comply with requests by optimizing user satisfaction.

LLM assistants thus generate texts that optimize for human preferences. This is misaligned with how a tutor should behave. Indeed, when a student asks for help, the tutor should not necessarily maximize the student's satisfaction, but rather their learning. For this, the tutor may have an intent (for example, asking back a question) different from what the student would like them to do (for example, giving them the solution). This mismatch is one of the core limitations of LLMs 
as tutors. %

\subsection{\SECllmsfortutoring{}}
Several works have attempted to address the lack of pedagogical properties of LLMs. The easiest and most direct approach is to write an extensive prompt~\cite{Bridge,SPOCK,autotutor_llm}. However, enumerating all desired behaviors of a tutor in words is complex, especially for nuanced pedagogical strategies.   

Another method considered is fine-tuning LLMs to efficient tutoring strategies. However, supervised fine-tuning is hampered by the scarcity of tutoring datasets, as their collection is costly and poses ethical concerns~\cite{GPTforGood}. \citet{mathdial} adopt an inverted Wizard-of-Oz approach by asking human teachers to tutor GPT-simulated students to create a comprehensive dataset of high-quality tutoring sessions and show that fine-tuning LLMs on this dataset improves their pedagogy. Similarly, the CLASS framework~\cite{SPOCK} uses an LLM to generate a synthetic conversational dataset with student-tutor interactions to fine-tune a smaller LLM. However, the quality of these conversations is not evaluated, and is prone to hallucinations.  Additionally, fine-tuning is expensive and restricts the LLM to the specific teaching style present in the training data, thus requiring multiple datasets for different pedagogical approaches. LearnLM~\cite{learnLM,LearnLM2} alleviates this issue by integrating pedagogical instruction following tuning, but still heavily relies on synthetic data and a costly co-training process where LearnLM is trained alongside a generic LLM.
The \texttt{Bridge} method~\cite{Bridge}, which we use as an inspiration for one of our baselines, uses an LLM to classify student errors and decide tutor interventions on the conversation history. Systems based on \texttt{Bridge} let the LLM choose the tutoring strategy on its own, \textit{assuming its ability to make pedagogically sound choices}. 

In contrast, our approach focused on letting learning scientists control the tutoring behavior. To do so, we employ a structured pedagogical intent selection using a transition graph based on classified student states, enabling experts to constrain the model to more faithfully follow a given pedagogy over multi-turn conversations.

\hlred{Recent work reconsiders the CLASS framework from the point of view of alignment and tackles it using common alignment techniques such as RLHF~\cite{pedagogically_aligned}. However, the data are synthetically generated which leads to questionable quality and bias, and as for fine-tuning, the need for data is the main drawback of this technique.}

\subsection{\SECpf{}}\label{section:PF}
In education, ``Instruction followed by Problem Solving'' approaches, which directly instruct students on how to approach a problem and then let them practice, are widely used. However, PF~\cite{productive_failure,Kapur_2016} rather uses the reversed order. PF is a ``Problem Solving followed by Instruction'' learning design, in which the learner attempts to solve open-ended problems before formally learning how to approach them, to build a deeper understanding of the topic. With PF, the teacher mostly uses questioning to let students engage cognitively with the problem.
There is a large body of evidence that PF is effective
with meta-analyses reporting an effect size of 0.87 standard deviations (high by the Learning Sciences standards)~\cite{meta_analysis_pf}, when compared to traditional teaching. %

A successful problem-solving session in PF provides opportunities for the learner to explore the problem and build as many Representation and Solution Methods (RSMs) as possible, even if they fail at solving the problem~\cite{Designing}. In simple terms, these RSMs can be seen as mental representations of the problem, or unique solution attempts. The number of RSMs generated is one of the best predictors of how much students learn from PF~\cite{nb_RSMs}. Our tutoring strategy thus aims to promote the generation of such RSMs, and we will use their number as one of the evaluation metrics.

Overall, PF is a suitable learning design to validate our approach because (1) it is a widely successful pedagogical strategy, and (2) opposed to native LLM behavior as a PF tutor should not directly reveal the solution to the students.

\section{\SECmethodology{}}\label{ourwork}

\subsection{\SECformalism{}}
\label{sec:formalism}
\paragraph{\SECdialogtutoringtask{}}
We extend the formalization of the Dialog Tutoring task defined in \cite{opportunities}. We consider a tutoring dialog history $\mathcal{H}_{\leq n} = (u_1, \ldots, u_n)$ made of tutoring turns $u_t = (y_t,x_t)$ with $y_t$ the tutor's utterance and $x_t$ the student's utterance at turn $t$. 
Each tutor utterance is associated with a set of \textit{tutoring intents} $\mathbf{i}_t$ whose elements are chosen from a taxonomy of intents $\mathcal{I} = \{I_1,\ldots,I_{|\mathcal{I}|}\}$. An \textit{intent} is defined as a goal of the tutor utterance, e.g. `\texttt{correct the student's mistake}'.
We denote by $K$ the grounding information, i.e. additional information provided to the LLM to give context. In our case, $K$ includes the tutoring exercise and its solution. The resulting generative conditional model produces output: 
\begin{equation}
y_{n+1} \sim p(\ y \ | \ \mathbf{i}_{n+1},\mathcal{H}_{\leq n},K)
\end{equation}

\paragraph{\SECtutoringstrategymodeling{}}\label{sec:tutoringstrategymodeling}
We define a 
\textit{tutoring strategy} 
function 
as a function $\phi$ that takes as input $(K,\mathcal{H}_{\leq t})$, the grounding information and dialog history up to any turn $t$, and that outputs a set $\mathbf{i}_{t+1} \subseteq \mathcal{I}$ of single-turn tutoring intents for turn $t+1$.
Abstractly, $\phi$ represents the decision process by which an expert tutor, given the context of the conversation, selects the best possible intents from a pedagogical standpoint to react to the student's last utterance. This function will be replaced in practice by an approximation $\tilde{\phi}$ introduced in the next paragraph. The resulting conditional tutor model generates $y_{n+1}$ according to: 
\begin{equation}
y_{n+1} \sim p( \ y \ | \ \phi(\mathcal{H}_{\leq n},K),\mathcal{H}_{\leq n},K)
\end{equation}

\paragraph{\SECstudentstatespace{} \label{sec:studentstatespace}}
Expert tutors base their interventions on observations about (1) the student's emotional or knowledge \textit{state}~\cite{motivation}, and (2) the state of the conversation and their previous interventions~\cite{efficient_techniques}. For example, a tutor does not react the same way to a computational mistake as to a reasoning one. Therefore, we use \textit{state features} (1) we define in the next paragraph, and the previous tutoring intents (2) as inputs to our intent selection function $\tilde{\phi}$.
\hlred{We propose the formalization developed in the next paragraph.}

Each student turn is associated with a set $\mathbf{s}_t$ of \textit{state features} describing the state of the student and of the conversation at turn $t$. The state features are chosen from a set $\mathcal{S}$ of possible features. A feature in $\mathcal{S}$ could, for example, be `\texttt{the student is confused}' or `\texttt{the student made a computational mistake}'. We denote $\psi$ a classifier that provides these features based on the dialog history and grounding. Our intent selection function thus writes $\widetilde{\phi}(\mathbf{i}_{n},\mathbf{s}_n) = \mathbf{i}_{n+1} \subseteq \mathcal{I}$. As previously noted, $\widetilde{\phi}$ depends on the student state features $\mathbf{s}_n$ and previous tutor intents $\mathbf{i}_{n}$.
Our final conditional generation model produces output:
\begin{equation}\label{eq:formalism}
    y_{n+1} \sim p(\ y \ |\ \widetilde{\phi}\left(\mathbf{i}_{n},\psi(\mathcal{H}_{\leq n},K)\right),
\mathcal{H}_{\leq n},
K)
\end{equation}

\begin{figure*}[h!]
    \centering
    \includegraphics[width=0.9\textwidth]{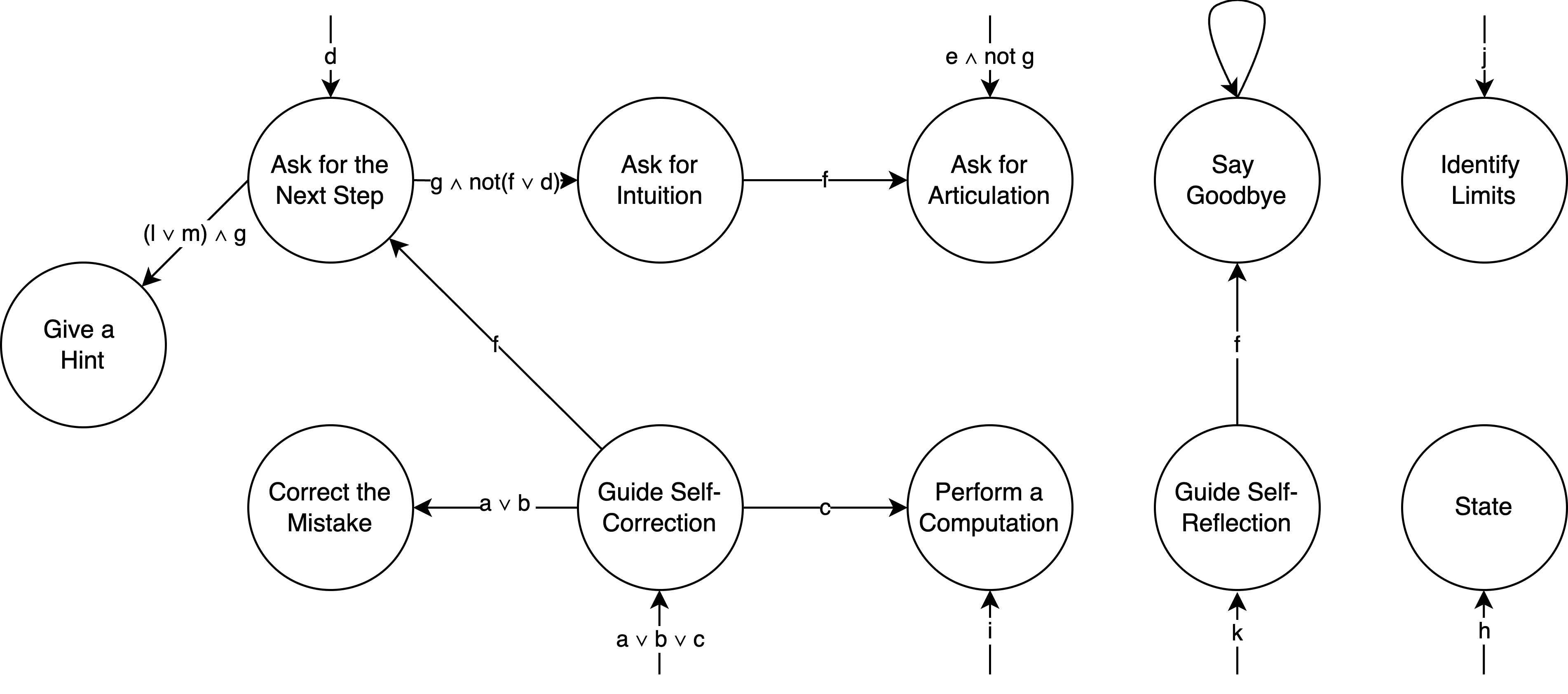}
    \caption{Transition Graph of a PF~\cite{productive_failure} Intent Selection. %
    Nodes correspond to intents (listed in Table \ref{tab:taxonomy} of Appendix \ref{sec:appendixtaxonomy}). The variables used in the arrows' transition conditions (`\texttt{a}', …, `\texttt{m}') are codes for the state features (displayed in Figure \ref{fig:Assessor_prompt} of Appendix \ref{sec:appendixassessor}). At each turn, we transition from one set of intents to the next by following all satisfied arrows. Arrows with no starting node can be taken from any node. At turn $0$, we start at node `\texttt{Ask for the Next Step}'. We describe the Learning Sciences justifications for this PF modeling in Appendix \ref{sec:appendixlearningsciencesbases}.} 
    \label{fig:transition_graph}
    \vspace{-0.8em}
\end{figure*}

\subsection{\SECalgorithm{}\label{sec:algorithm}}
\hlred{The Pedagogical Steering problem is caused by a misalignment between an LLM's goal and tutoring goals, and by LLMs' single- rather than multi-turn dimension.}
To steer the LLM towards a multi-turn pedagogical strategy, we introduce \name, an algorithm for the formalization (\ref{eq:formalism}) given in Section~\ref{sec:formalism}. \name\ dynamically updates the tutoring intents after every student utterance and prompts the LLM to follow them at the single-turn level. A visual representation of \name\ is presented in Figure~\ref{fig:main}. 
This algorithm is based on three sub-procedures: I. Student State Tracing, II. Tutor Intent Selection and III. Intent Dependent Steering.

\textit{I. The Student State Tracing} mirrors the function $\psi$. This step is a multi-label classification of the dialog history and latest student utterance at turn $t$. The selected features $\mathbf{s}_n$ are chosen from a pre-defined set of possible features $\mathcal{S}$. These features fall into four categories: (1) mistake types, (2) student requests, (3) conversation states, and (4) emotional states. These categories each provide valuable information to be used in the Intent Selection, i.e.: (1) helps plan error remediation (e.g. a reasoning and a computational mistake should not be dealt with equally)~\cite{stepwise,Bridge}, (2) is necessary as different student requests call for different reactions (e.g. if the student asks for the solution v.s. if they ask for a definition), (3) helps structure the conversation (e.g. knowing if the problem is solved) and (4) allow for proper emotional responses from the tutor~\cite{motivation}. The precise features we use are detailed in Appendix~\ref{sec:appendixassessor}. State tracing is performed by an LLM prompted to identify these different features given the set of possible student states $\mathcal{S}$ and the dialog history. Implementation details such as the prompt we used are provided in Appendix~\ref{sec:appendixassessor}.

\textit{II. The Tutor Intent Selection} mirrors the function $\widetilde{\phi}$. We model tutor intent selection as a deterministic transition from the previous set of intents $\mathbf{i}_n$ to the next set of intents $\mathbf{i}_{n+1}$, conditioned on the output $\mathbf{s}_n$ of step I. The transition rules are defined by a learning scientist expert. \hlred{, but future work could learn them from annotated data.} We represent these transitions with a graph. Please note that, unlike traditional deterministic state transition models, our graph allows to transition from a \textit{set} of nodes to another by following multiple concurrent transitions.

We used the theories of Scaffolding~\cite{structuring_problematizing_scaffolding}, Affective Cognition~\cite{motivation}, and PF~\cite{when_pf_fails,Designing} to design a taxonomy of tutoring intents $\mathcal{I}$. These intents are mainly separated into: (1) \textit{scaffolding intents}, helping the student by restricting the solution space (e.g. giving a hint), and (2) \textit{problematizing intents}, helping the student deepen their understanding by exploring the solution space (e.g. asking to formalize an intuition). For our PF strategy, we prioritize scaffolding when the student is stuck and use problematizing intents whenever the student shows an understanding of the problem, to encourage the free exploration of the problem and the generation of multiple RSMs. The exhaustive taxonomy is displayed in Table~\ref{tab:taxonomy} of Appendix~\ref{sec:appendixtaxonomy} and a detailed explanation of the Learning Science principles we used to design our PF Intent Selection is given in Appendix \ref{sec:appendixlearningsciencesbases}.

Our PF Intent Selection graph is represented in Figure~\ref{fig:transition_graph}. Each node represents an intent of $\mathcal{I}$. Arrows correspond to transitions between intents. We initialize start at the `\texttt{Ask for the Next Step}' node and move at each tutoring turn. The letters used in the logical conditions on the arrows are codes corresponding to Student State Tracing features (detailed in Figure~\ref{fig:Assessor_prompt} of Appendix \ref{sec:appendixassessor}).
For the sake of clarity, transitions that can be taken from any intent are represented by arrows with no starting node.

\textit{III. The Intent Dependent Steering} generates the LLM Tutor prompt $p_{n+1}$ conditioned on the selected tutoring intents $\mathbf{i}_{n+1}$. We defined a one-to-one correspondence between the intents in $\mathcal{I}$ and short prompt chunks. The precise correspondence is given in Appendix \ref{sec:appendixpromptgenerator}. \hlred{we call prompt additions. As depicted in Figure~\ref{fig:main}, the generated prompt $p_{n+1}$ is made of a base prompt template, followed by the grounding information, and an intent-specific prompt addition for each intent in $\mathbf{i}_{n+1}$.}

\section{\SECexperiments{}}\label{assessment}
\textbf{Experimental Setup. }
As a case study, we use \name\ to implement a prototype high school math tutor based on GPT-4o (gpt-4o-2024-08-06) and following a PF--inspired tutoring strategy. Our tutor discusses one problem per tutoring dialog and has access to its solution. The exact prompts we used in our prototype, and the exhaustive list of state features, intent taxonomy, and intent-to-prompt correspondences are given in Appendix \ref{sec:appendiximplementation}.

\subsection{\SECrqs{}}
We validate the ability of our system to control the tutoring strategy employed by the LLM by assessing the capacity of our prototype to follow the PF paradigm.\hlred{We want our system to do so} Ideally, without hindering the factual correctness or desirable conversational properties of LLMs such as their coherence. We formulate these goals in two research questions:

\begin{enumerate}
    \item[]{\textbf{RQ1.}}\label{RQ1} To what extent does \name\ make the LLM Tutor follow PF?
    \item[]{\textbf{RQ2.}}\label{RQ2} Does \name\ influence on other desirable properties of the LLM Tutor?
    \hlred{
    \begin{enumerate}
        \item[]{\textbf{RQ2.1.}}\label{RQ2.1} Does \name\ influence the ability of the LLM to generate factually correct responses?
        \item[]{\textbf{RQ2.2.}}\label{RQ2.2} Does \name\ influence the ability of the LLM to act like a human?
        \item[]{\textbf{RQ2.3.}}\label{RQ2.3} Does \name\ influence the perceived usefulness of the tutor?
    \end{enumerate}
    }
\end{enumerate}
We chose to handcraft the Intent Selection mechanism with the help of PF experts. \hlred{, assuming that an LLM could not model such a complex strategy by choosing intents on its own.} This is different from~\citet{Bridge} which lets an LLM choose how to react. We thus formulate a design question to test our assumption: \dqone{}.\label{DQ1} How does the intent selection by LLM compare to our approach?

\subsection{\SECdataset{}\label{sec:dataset}}

We compiled a dataset of \nproblems\ PF problems. These problems are open-ended, ill-structured, and engaging. They feature multiple suboptimal solutions (and a single optimal one) and require multiple interactions from students to explore their solution space. These serve as high-quality testing problems for evaluating a tutoring system’s pedagogical skills.
Our algorithm is tested on two 9th-grade math problems (successfully solved by 62\% of our students in the field test). The first problem, \texttt{Consistency}, is adapted from~\citet{consistency_pb} and is a well-studied \textit{invention problem} where students devise a consistency measure (typically variance) to compare football players' performance. An effective PF tutor for this problem would help the students construct as many ways to measure consistency as possible. The second problem, entitled \texttt{Country}, is an open-ended problem taken from~\citet{coutry_pb} requiring students to fairly allocate voting seats across states based on their population. On this problem, an effective PF tutor would guide reasoning without explicitly providing key steps like the idea of using proportionality. The exact problem statements, solutions, and the dataset specifications, intended use, and license are provided in Appendices \ref{sec:appendixdataset} and \ref{sec:appendixtestproblems}.

\subsection{\SECbaselines{}}
We test our \name-controlled tutor LLM (\textbf{V1}) against three ablation versions.
\textbf{V2: No Intent} is a baseline LLM prompted to act like a tutor with our base prompt (but no intent-dependent prompt).
\hlred{We use it to assess the tutoring strategy fidelity of our system.}
\textbf{V3: Constant Intent} always uses the same intent (``Ask for Next Step'' in Table \ref{tab:taxonomy} of Appendix \ref{sec:appendixtaxonomy}) that only asks the student to proceed with the exercise. We use this baseline to assess the importance of the Intent Selection procedure.
\textbf{V4: LLM Intent} use an LLM to perform the Intent Selection (with the prompt given in Appendix \ref{sec:appendixintentselectorprompt})~\citet{Bridge}, 
and we use it to answer \hlred{whether the intent's choice needs to be determined by an expert-designed procedure or if a natural language description of the tutoring strategy to follow is enough (}\dqone{}.  

\subsection{\SECevaluationmetrics{}}\label{sec:evaluationmetrics}
To quantitatively assess the \textit{PF fidelity} of the different versions (their success in following PF), we have to rely on proxy variables.
We propose as a main proxy the \textbf{number of student-generated Representation and Solution Methods} (RSMs), which can be seen as the different ways students represent the problem and try to solve it. As explained in Section \ref{section:PF}, the diversity of student-generated RSMs is the first predictor of learning gains with PF~\cite{Designing,nb_RSMs}. We thus use this number to measure the PF fidelity of the tutor. 

However, not every problem easily admits multiple RSMs. In our case, only \texttt{Consistency} does. We thus define another proxy, the \textbf{PF Score}. One of the core ideas of PF is to encourage the student to construct their own understanding of the problem by exploring it by themselves, even if they do not succeed in solving it. Building on this idea, we define for each problem a grading rubric used to compute the tutor's PF Score: For each problem, we identify a list of critical reasoning steps that represent different conceptual understandings of the problem. To assess a tutoring conversation, we give a point to the tutor for every crucial reasoning step that the student found on their own. We do not award the point if the tutor reveals it or directly hints to it. This metric can be thought of as an adjusted sub-step success rate. An efficient PF tutor would help the student explore the problem and thus achieve these conceptual understandings, without giving them any direct hint or revealing them, resulting in a high PF Score. We display the precise sub-steps for both testing problems in Appendix \ref{sec:appendixpfscoredetails}.

\subsection{\SECsyntheticstudy{}}

Due to ethical concerns~\cite{GPTforGood}, we first tested our prototype with GPT-4–simulated students before involving humans. This method of using simulated students, known as interactive tutoring, is used in the literature~\cite{SPOCK,mathdial}. Importantly, we do not assume that the LLM can accurately model students' learning. However, this assumption is not necessary for our preliminary study as we are interested in the ability of our algorithm to follow a PF tutoring strategy, rather than the learning gains it produces. We simulated three conversations for each of two problems across four tutor versions (V1--4) and used a scoring rubric to annotate conversations, computing the PF Score and RSM count (as described in Section \ref{sec:evaluationmetrics}).

\subsection{\SECfieldtest{}}
We ran the field test with 17 9th-grade students from a high school in Singapore. Each student was randomly (uniformly) assigned to one of two conditions (V1 or V2, chosen as described in Section~\ref{res_synthetic}) and one of the two exercises. They had 15 minutes to interact with the tutor. We manually annotated their conversations using our rubric to compute the PF score and the number of generated RSMs.

\textbf{\SECfeedbackquestionnaire{}. }
After the tutoring session, students were asked about their experience to assess potential spillover effects of StratL on the tutor LLM’s desirable properties. Using a 3-point Likert scale, they rated the AI tutor on: a) coherence, b) empathy, and c) helpfulness.

\section{\SECresultsdiscussion{}}\label{results}
\subsection{\SECresultssyntheticstudy{}}\label{res_synthetic}
\begin{table}[h!]
    \centering
    \caption{PF fidelity metrics for different tutor versions and different problems in the preliminary study with simulated students. Best values in bold. 
    }
    \label{tab:PF_fidelity_evaluation}
    \resizebox{\linewidth}{!}{
    \begin{tabular}{lccc}
      \toprule
      & \textbf{\# RSMs} & \multicolumn{2}{c}{\textbf{PF Score}} \\
      \cmidrule(lr){3-4}
      & & \texttt{Consistency} & \texttt{Country} \\
      \midrule
      \textbf{V1. \name} & \textbf{2.34} & 3.67 & \textbf{4.00} \\
      \textbf{V2. No Int.} & 1 & 1.33 & 0.44 \\
      \textbf{V3. Constant Int.} & 1.34 & \textbf{4.00} & \textbf{4.00} \\
      \textbf{V4. LLM Int.} & 1.34 & 3.67 & 3.11 \\
      \bottomrule
\end{tabular}}
  \end{table}

\textbf{\dqone{}.}
\emph{Using an LLM prompted with a natural language description of the tutoring strategy is not enough for the Intent Selection.}
The preliminary study in Table \ref{tab:PF_fidelity_evaluation} suggests that \name\ (V1) is successful in steering the LLM tutor towards PF, as evidenced by the greater number of RSMs elicited by this version. We observe that the version using \name\ with an LLM responsible for the Intent Selection (V4) did perform better than the baseline LLM (V2), but not as well as \name\ with our handcrafted Intent Selection mechanism. This result contributes to an answer to \dqone{} by showing that providing a natural language description of the tutoring strategy to follow and letting an LLM choose the intents accordingly does not yield satisfactory results in terms of tutoring strategy fidelity.

\textbf{RQ1.}
\emph{The strategy followed by an LLM controlled by \name\ is determined by the Intent Selection procedure.}
Overall, V3 achieves a rather high PF Score on the two problems, since it never provides any hint to students. However, this result is more a consequence of the way we defined the PF score rather than a sign of PF fidelity, as V3 does not perform as well on our second proxy: It does not manage to elicit as many RSMs from students as V1, with an average of only 1.34. This shows that the strategy followed by an LLM controlled by \name\ is determined by the Intent Selection procedure as it is the only part that differs (V1 vs. V3).

Based on our preliminary answers to \textbf{DQ1} and \textbf{RQ1}, \textbf{we decided not to include V3 and V4 in our field test} with real students to focus on the best-performing version (V1) while keeping the non-\name\ baseline (V2) for comparison.

\textbf{\SECaccuracystudentstatetracing}\label{sec:accuracystudentstatetracing}
The functioning of \name\ as a whole is contingent upon the functioning of the Student State Tracing.
To test its reliability, we generate ten tutoring conversations for both problems and the first author manually annotates them with ground truth state features $\mathbf{s}_t$. We 
compare the outputs of the Student State Tracing with the annotated data.
As a multi-label classification task, we compute for each state feature the precision, recall, and F1 score. 
We compare three different implementations of the State Tracing step and provide more details about our methodology and results in Table \ref{tab:Assessor_test} of Appendix \ref{sec:appendixstudentstatetracingtes}.
Our implementation achieves a satisfactory micro F1 score of $0.77$.
The Student State Tracing is a task of its own, and we encourage researchers to try different classification models to improve its reliability.

\subsection{\SECresultsfieldtest{}}
\begin{figure}
    \begin{center}
        \resizebox{\linewidth}{!}%
        {\includegraphics{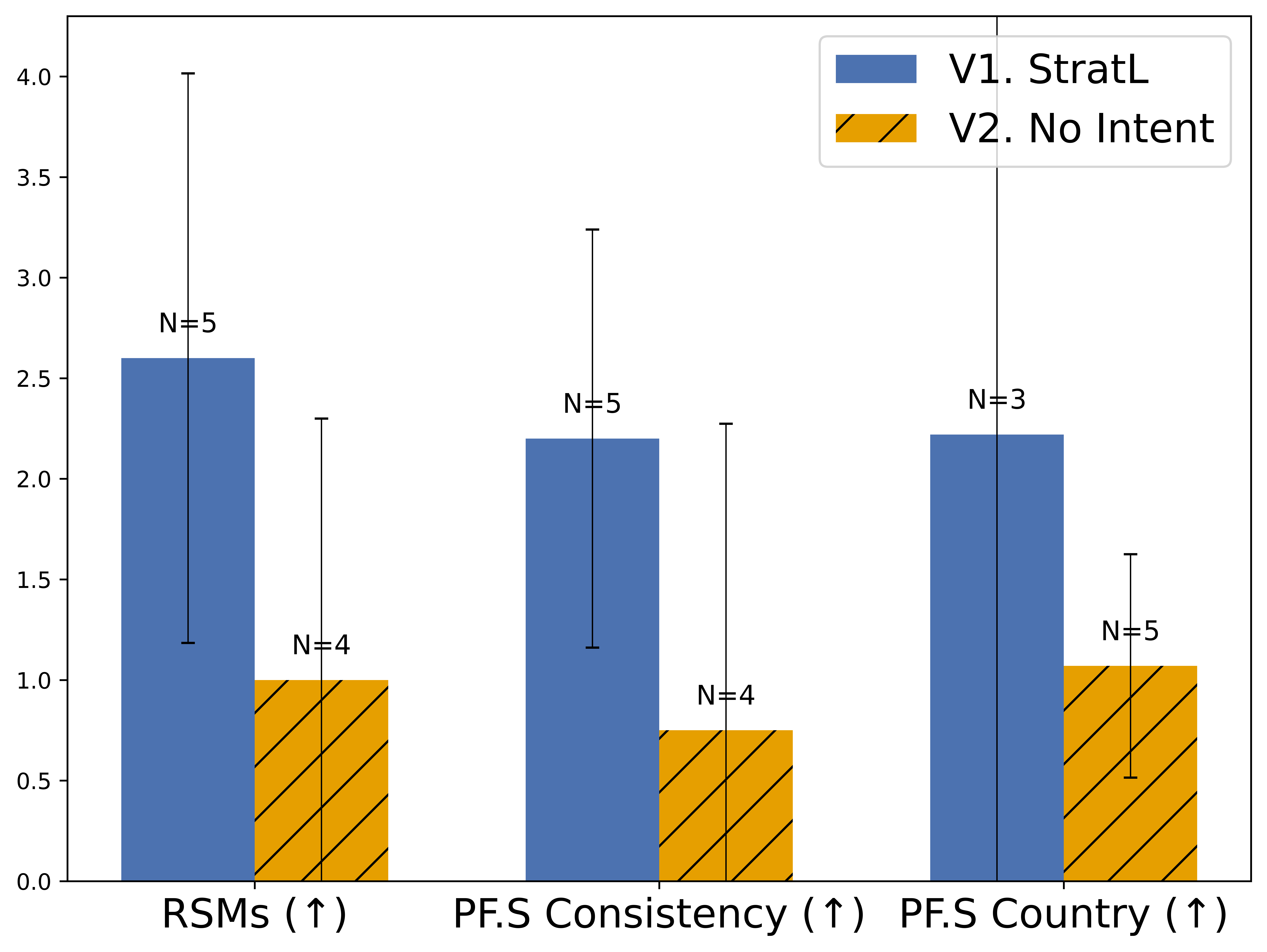}}
    \end{center}
    \vspace{-1.2em}
    \caption{Field test results and their confidence intervals. 
    `PF.S' refers to the `PF Score' as defined in Section \ref{sec:evaluationmetrics}{} and is out of 4. `RSMs' denotes the number of student-generated RSMs. 
    \name\ (V1) succeeds in steering the LLM to follow PF as compared to V2.}
    \label{fig:study_results} 
     \vspace{-.8em}
\end{figure}

\begin{table}[h!]
  \centering
  \caption{Feedback questionnaire results (out of 3). `N' refers to the number of answers to the questionnaire. We do not detect any harmful effect of \name\ (V1) on the LLM tutor's coherence, empathy, or perceived helpfulness.}
   \label{tab:human_evaluation}
\begin{tabular}{lcccc}
    \toprule
    & \textbf{N} & \textbf{Coherence} & \textbf{Empathy} & \textbf{Helpful.} \\
    \midrule
    \multicolumn{5}{l}{\textit{Problem: Consistency}} \\
    \midrule
    \textbf{V1.} & 5 & 1.80 & 1.60 & 1.20 \\
    \textbf{V2} & 4 & \textbf{2.25} & \textbf{1.75} & \textbf{2.00} \\
    \midrule
    \multicolumn{5}{l}{\textit{Problem: Country}} \\
    \midrule
    \textbf{V1} & 3 & \textbf{3.00} & \textbf{2.67} & 2.67 \\
    \textbf{V2.} & 5 & \textbf{3.00} & 2.20 & \textbf{2.80} \\
    \bottomrule
\end{tabular}
\vspace{-0.8em}
\end{table}

\textbf{\SECresultsfidelity{}}\label{pf_fidelity}
\emph{\name\ is successful in steering the LLM to follow the PF tutoring strategy.}
We report in Figure \ref{fig:study_results} the results of the protocol defined in Section \ref{sec:evaluationmetrics}{}. We note the PF score (out of 4) is used as a proxy for the extent to which the tutor followed the PF tutoring strategy. We also report the average number of student-generated RSMs (partial solutions) on the problem \texttt{Consistency}. This metric is best proxy for PF fidelity, as one of the goals of the PF tutoring strategy is to maximize this number, as explained in Section \ref{section:PF}. Out of the two test problems, the number of RSMs can only be measured on \texttt{Consistency} as it is the only invention problem.

On both test problems, the \name-controlled LLM (V1) achieved a higher (statistically significant for problem \texttt{Consistency}, $p = .046$ with a 2-sample t-test) PF score than the baseline LLM (V2), signifying a higher fidelity to the PF tutoring strategy. Furthermore, the baseline LLM (V2) failed to elicit multiple RSMs from students, while a student tutored by the \name-controlled LLM produced on average 2.6 RSMs during the tutoring session (the difference is statistically significant, $p = .05$ with a 2-sample t-test). In that regard, V1 succeeded in the PF process of making students explore as many solutions as possible.

\textbf{\SECresultsfeedback{}}
The feedback questionnaire did not reveal particular spillover effects of \name\ on the coherence, empathy or perceived helpfulness of the tutor LLM.

\emph{\name\ has no or little influence on the capacity of the tutor LLM to generate coherent responses.} 
\name\ (V1) scores similarly or slightly less than an LLM with no intent (V2) on coherence across the two problems: $1.80$ out of $3$ for (V1) \textit{vs.} $2.25$ for (V2) when used on problem \texttt{Consistency}, and $3.00$ for both on \texttt{Country}. 
We do not detect any statistically significant difference.

\emph{\name\ does not influence the capacity of the LLM to generate empathetic answers.} 
None of the two versions is perceived as significantly more empathetic than the other ($1.60$ for V1 \textit{vs.} $1.75$ for V2 on \texttt{Consistency} and $2.67$ \textit{vs.} $2.20$ on \texttt{Country}). %

\emph{\name\ can influence the perceived helpfulness of the tutor LLM, through the tutoring strategy it makes it follow.}
We find that the baseline with no intents (V2) is perceived by the students as more helpful than V1 ($2.00$ \textit{versus} $1.20$ respectively on \texttt{Consistency} and $2.80$ \textit{versus} $2.67$ on \texttt{Country}). %
This is not surprising, as the PF tutoring strategy can be frustrating or misunderstood by students. Indeed, the PF strategy is designed to only give help when it is strictly necessary, which means that students spend more time being stuck and exploring suboptimal solutions. This can lead students to perceive the tutor as less helpful compared to a version that would give more help. We report the detailed results in Table \ref{tab:human_evaluation}.

\subsection{\SECdiscussions{}}
Our field test demonstrated the ability of the \name\ tutor to follow PF. This section highlights some qualitative insights from the field study feedback.

\textbf{Robustness}
As discussed in Section \ref{sec:accuracystudentstatetracing}, the Student State Tracing procedure can make labeling mistakes. These mistakes can then compromise the Intent Selection and result in the selection of contradictory intents. This is exemplified by the feedback of a student: \textit{``The AI keeps on asking me to check my answer although it was correct''}.

\textbf{Solution constraint}
A student assigned to the no intents baseline (V2) complained that the tutor would not let them use a solution entirely different from the expected one: \textit{``The AI only listens to key points that it wants, and ignores everything else (e.g. my idea of using [...])''}
A good tutor should encourage original solutions. This highlights a pedagogical problem, as in this case, the tutor is harmful to the learning by restricting the student.

\section{\SECconclusionslimitations{}}\label{conclusion}
We present the pedagogical limitations of LLMs in tutoring as the Pedagogical Steering problem. We introduce \name, an algorithm addressing this problem. %
\name\ integrates ideas from Learning Sciences with LLMs, utilizing a Student State Tracing and a Tutoring Intent Selection procedure based on expert-defined transition graphs to dynamically steer the LLM to follow a modeled pedagogy. \hlred{We use productive failure learning design in the domain of high school math to test \name.} A field study with high school students demonstrates \name's ability to follow the Productive Failure learning design. %

\section{Limitations}

\textbf{Evaluation aspects.} The field study uses a rubric-based evaluation and learner-perceived usefulness in an end-to-end scenario. We argue additional evaluation metrics such as multiple human teachers evaluating the quality across the dimensions such as engaging the student or quality of questioning. Moreover, as we evaluated \name\ through a field study, learners might have taken the exercise less seriously than they would for an exercise that is officially part of their curriculum. Further studies with increased ecological validity should be conducted.

\textbf{Integration. } While tutoring can also be explored as a standalone experience, it is important to consider how systems like \name\ could be integrated into a classroom or as part of a formal education context. As such, further research should explore (1) how to support teachers in adjusting and personalizing the tutor's strategy, (2) how to provide relevant feedback to the teacher in order to ensure continuity with classroom content, and (3) how such tutors should be best integrated in existing pedagogical structures.

\textbf{Risks.}
As students increasingly utilize AI tools in their learning process, we must ensure they are beneficial to learning. Some studies empirically demonstrated that generative AI tools could harm learning~\cite{genaiharmlearning}, through mechanisms such as overreliance~\cite{overreliance}. With our work, we mitigate these effects by empowering teachers and learning scientists to control their student's interactions with generative AI through the use of \name. However, further field studies on the reliance behavior of students with a \name LLM tutor should be conducted before large-scale deployment. \citet{survey} discusses more broadly the risks associated with the use of LLMs in education.

\textbf{Social aspects.} The social context of a learning experience plays an important role in learning, and ``social surround'' is even one of the three design layers of PF~\cite{Designing}. Generally, supportive and coordinated relationships play an important role in learning~\cite{roschelle1995construction}, and research on how students perceive these tutors at the social level should be explored to paint a complete picture. 
Beyond cognitive learning outcomes, it is also important to consider affective outcomes of learning activities, such as social-emotional competencies~\cite{alzahrani2019effect}, and how these are impacted by the interaction with digital tutors over human ones.

\textbf{Data Privacy.} The field study was approved by the university IRB.
All data collected were anonymized and no sensitive student information was stored.

\bibliography{main}

\appendix

\begin{table*}[ht]
\footnotesize
\begin{tabular}{|p{3.5cm}|p{2cm}|p{2.5cm}|p{2cm}|p{2cm}|p{1.5cm}|}
\hline
\textbf{type} & \textbf{grade} & \textbf{name} & \textbf{reference} & \textbf{exercise} & \textbf{solution} \\ \hline
Type between \texttt{invention}, \texttt{ill-structured}, and \texttt{well-structured} & Recommended school grade & Name to quickly identify the exercise & Article the exercise was curated from & Exercise statement & Solution if provided in article \\ \hline
\end{tabular}
\caption{Description of the fields in the dataset.}
\label{tab:dataset_fields}
\end{table*}

\section{Appendix Evaluation Details\label{sec:appendix_eval}}

\subsection{\SECappendixdataset{}\label{sec:appendixdataset}}

We collected \nproblems\ problems from the PF literature  ~\cite{productive_failure,Designing,PF_in_math,PFcollaboration,coutry_pb,consistency_pb,loibl,consistency_tea,comparing_fractions_2}.
We present the columns of the dataset in Table~\ref{tab:dataset_fields}. We curated three types of problems: \textit{Invention} problems ask the student to invent a new way to solve a problem. They typically admit several correct but suboptimal answers, as long as the student can justify them. Students are not expected to have previous exposure to the tools canonically used to solve the exercise. The problem \texttt{Consistency} is a good example of \textit{invention} problem. \textit{Ill-structured} problems are open-ended problems that contain some non-relevant information and a convoluted story. These problems often admit a single solution, but there can be several ways to reach it. The problem \texttt{Country} is an example of such a problem. \textit{Well-structured} problems are more direct versions of the \textit{ill-structured} problems. They resemble classical exercises that students usually solve in a mathematics class. More information about the different types of exercise in PF and their role can be found in \cite{productive_failure} and \cite{Designing}.
Our dataset is intended to be used as a set of high-quality testing problems for evaluating a tutoring system’s pedagogical skills. We release it under CC-BY-4.0 license\footnote{\url{https://creativecommons.org/licenses/by/4.0/deed.en}}.

\subsection{\SECtestproblems{}}\label{sec:appendixtestproblems}
Figures~\ref{ex:country} and \ref{ex:consistency} display the two test problems we used in our field test. They are also included in the dataset we release.

\subsection{PF Score and RSM Details}\label{sec:appendixpfscoredetails}
\paragraph{PF Score} We present in the following paragraphs the precise reasoning sub-steps used for the PF Score grading as explained in Section \ref{sec:evaluationmetrics}{}.

For problem \texttt{Consistency}, we define these reasoning steps as 1. Giving an intuitive definition of consistency related to the variation of scores, 2. Thinking about comparing the scores to a baseline (central value or previous year), 3. Considering absolute differences (by taking the absolute value or squaring) since only the magnitude of the change matters, and 4. Thinking about a way of aggregating all the differences into a single score, for example by summing or averaging them. 

For problem \texttt{Country}, we define these steps as 1. Thinking about using a `standard divisor' which represents the number of seats per inhabitant, 2. Thinking about rounding down the number of seats not to attribute more seats than the total, and 3. Thinking about using the remainders to distribute the additional seats. We linearly scale the PF score to 4 in this case.

\paragraph{RSM} We tracked the number of student-generated Representation and Solution Methods (RSMs) in problem \texttt{Consistency}. We observed students proposing various solutions to the problem, including comparing the three soccer players based on the variance of their number of goals scored or on their mean absolute year-to-year differences.

\begin{figure}[!htb]
  \begin{framed}
  \scriptsize
A new country has recently been founded.
The country is split into six states, call them A, B, C, D, E, and F. \newline
The population of state A is 1,646,000 people, the population of state B is 6,936,000 people, the population of state C is 154,000 people, the population of state D is 2,091,000 people, the population of state E is 685,000 people, and the population of state F is 988,000 people.\newline
There are 250 seats available on a legislative body to govern the new country. 
How many seats should be assigned to each state so that each state would receive a fair representation? \newline
Show your work and justify why you think your method is correct.\newline

\underline{Solution}:\newline
We assign sits proportionally to the population of each state. Since the results of the divisions are not integers, we round down the number and then distribute the remaining sits to stats having the largest remainders.\newline

Total population = 1,646,000 (A) + 6,936,000 (B) + 154,000 (C) + 2,091,000 (D) + 685,000 (E) + 988,000 (F)      
                  = 12,500,000\newline

Standard divisor = Total population / Number of seats = 12,500,000 / 250
                  = 50,000\newline

Initial quotas:\newline
   - A: 1,646,000 / 50,000 = 32.92 → 32 seats\newline
   - B: 6,936,000 / 50,000 = 138.72 → 138 seats\newline
   - C: 154,000 / 50,000 = 3.08 → 3 seats\newline
   - D: 2,091,000 / 50,000 = 41.82 → 41 seats\newline
   - E: 685,000 / 50,000 = 13.70 → 13 seats\newline
   - F: 988,000 / 50,000 = 19.76 → 19 seats\newline

   Total initial seats assigned = 32 + 138 + 3 + 41 + 13 + 19 = 246\newline

   Seats left to distribute = 250 - 246  = 4\newline

Distribute the surplus seats based on largest remainders:\newline

   Remainders (from the divisions above):\newline
   - A: 0.92\newline
   - B: 0.72\newline
   - C: 0.08\newline
   - D: 0.82\newline
   - E: 0.70\newline
   - F: 0.76\newline

   The four highest remainders are from states A, B, D, and F. Give one extra seat to each.\newline
  \end{framed}
  \caption{Problem \texttt{Country} and its solution.}
  \label{ex:country}
  \vspace{-10pt}
\end{figure}

\begin{figure}[!htb]
  \begin{framed}
  \scriptsize
  The organizers of the Premier League Federation have to decide which one of the three players Mike Arwen, Dave Backhand and Ivan Right - should receive the "The Most Consistent Player for the Past 5 Years" award. Table 1 shows the number of goals that each striker scored between 2019 and 2023.

The organizers agreed to approach this decision mathematically by designing a measure of consistency. They decided to get your help. Here is what you must do:\newline
(1) Design as many different measures of consistency as you can.\newline
(2) Your measure of consistency should make use of all data points in the table.\newline

{
Table 1. Number of goals scored by the three players in the Premier League between 2019 and 2023.

\begin{tabular}{|c|c|c|c|}
\hline
Year & Mike Arwen & Dave Backhand & Ivan Right \\
\hline
2007 & 13 & 12 & 14 \\
2008 & 12 & 14 & 10 \\
2009 & 15 & 16 & 18 \\
2010 & 17 & 15 & 18 \\
2011 & 13 & 13 & 15 \\
\hline
\end{tabular}\bigskip
}

\underline{Solution}:\newline
The concept of variance and standard deviation is unknown to students.
Any measure proposed by the student is acceptable as long as it can be justified to measure consistency.
The goal is for them to construct their own measure of consistency and justify it based on the data provided.

Example of canonical solution: computing the variance (or standard deviation) for each player (standard deviation is also valid):\newline

First, compute the mean:\newline
Mean number of goals for Mike: 14\newline
Mean number of goals for Dave: 14\newline
Mean number of goals for Ivan: 15\newline

Then, compute the sum of square deviations from the mean for each player:\newline
Sum squared deviation for Mike: 16\newline
Sum squared deviation for Dave: 10\newline
Sum squared deviation for Ivan: 44\newline

Then divide by the number of data points to get the variance:\newline
Variance for Mike: 12/5 = 3.2\newline
Variance for Dave: 10/5 = 2\newline
Variance for Ivan: 44/5 = 8.8\newline

So according to the variance, Dave is the most consistent player.
  \end{framed}
  \caption{Problem \texttt{Consistency} and its solution.}
  \label{ex:consistency}
  \vspace{-10pt}
\end{figure}

\subsection{\SECappendixstudentstatetracingtest\label{sec:appendixstudentstatetracingtes}}
As explained in Section \ref{sec:accuracystudentstatetracing}, we evaluated the accuracy of our Student State Tracing procedure, including a comparison against baselines.

\begin{table}[htbp]
  \centering
  \caption{Test micro-averages for different versions of the Student State Tracing.}
  \label{tab:Assessor_test}
  \resizebox{\columnwidth}{!}{
  \begin{tabular}{lcccc}
    \toprule
    & $\mu$-average precision & $\mu$-average recall & $\mu$-average F1 \\
    \midrule
    \textbf{Full version} & \textbf{0.82} & \textbf{0.75} & \textbf{0.77}\\
    \textbf{No Justification} & 0.71 & 0.72 & 0.70 \\
    \textbf{Short Memory} & 0.64 & \textbf{0.75} & 0.68 \\
    \bottomrule
  \end{tabular}}
\end{table}

\paragraph{Methodology}
We test the procedure using two tutoring conversations of about ten dialog turns for each of the two test problems detailed in Appendix \ref{sec:appendixtestproblems}. These conversations are manually annotated to obtain the reference data. As detailed in Appendix \ref{sec:appendiximplementation}, we use an LLM to perform this task in our implementation of \name. This LLM is prompted with the full conversation history and is asked to output a justification for the labels it selected. We compare this implementation to a version not outputting justifications and a version that only keeps the three last utterances in the prompts' history to increase the processing speed.

As a multi-label classification task, we compute for each label the precision, recall, and F1 score. We aggregate these metrics across labels using an average weighted by the relative cardinality of each label, known as the micro average ($\mu$-average). We report the results in Table \ref{tab:Assessor_test}.
\paragraph{Results} Our simple Student State Tracing already achieves satisfactory results. The use of justification indeed seems to increase the State Tracing's performance as we can observe a higher precision and F1 score for the version using it compared to the version not justifying its answers. It is worth remembering that this increased performance comes at a higher cost, since the LLM has to generate additional tokens for the justification. The State Tracing model keeping in memory only the last three dialog pairs does not perform as well as the full version: it thus appears important to provide the whole context of the conversation to obtain reliable results.

\subsection{Instructions given to the Field Test Participants}

Figure \ref{instructions} displays the instructions and consent form given to the field test participants before entering the study.

\begin{figure*}[t]
  \begin{framed}
  \scriptsize
Participant ID: <ID>. \textbf{Please write down and keep this ID for future reference.}  \\
Contact project team: <REDACTED>   \\
Data Protection Officer: <REDACTED>    \\
\newline
\textbf{We would like to ask you if you are willing to participate in our research project. Your participation is voluntary. Please read the text below carefully and ask the contact person about anything you do not understand or would like to know.} \\
\newline
\#\#\#\#\# What is investigated and how? \\
You will have the opportunity to test our AI math tutor. We will investigate whether our AI tutor is able to follow a given pedagogical strategy. You will be asked to solve a math exercise with the help of the tutor. We will analyze your tutoring session to extract some numerical metrics related to how good the tutor was. The results will be aggregated into statistics. You will then give some feedback on the tutor. \\
\newline
\#\#\#\#\# Who can participate? \\
You need to be able to write early high school-level mathematics in English to participate.\\
\newline
\#\#\#\#\# What am I supposed to do as a participant?\\
As a participant, you will connect to the website provided by your teacher. You will find a chat interface with an AI math tutor on this website. You will first read the AI tutorial, and then chat by text with the AI tutor to solve the exercise. After 15 minutes with the tutor, you will then fill in a feedback form on your experience with the tutor.\\
\newline
\#\#\#\#\# What are my rights during participation?\\
Your participation in this study is voluntary. You may withdraw your participation at any time without specifying reasons and without any disadvantages.\\
\newline

\#\#\#\#\# What risks and benefits can I expect?\\
We don't expect any risk for you. Your conversation with the tutor will be recorded in an anonymous fashion and won't be shared with anyone beyond the researchers involved in the project. The AI tutor might sometimes make mistakes. The tutor is programmed to follow a pedagogy that aims to improve your understanding of math. This session can thus be a very good practice session for math.\\
\newline

\#\#\#\#\# Will I be compensated for participating?\\
You will not be compensated for participating in this study. This educational activity will help you learn math.\\
\newline

\#\#\#\#\# What data is collected from me and how is it used?\\
Your tutoring session (the messages you send to the AI and the ones you receive) will be recorded. The post-test feedback form will ask you your gender, if you like math, and will ask you to rate your experience with the AI tutor. Only the authors will have access to this data, and it will be anonymous.\\
Members of the <REDACTED> may access the original data for examination purposes. Strict confidentiality will be observed at any time.\\
Results will be published in conferences in a completely anonymous fashion. The conversations with the tutor will not be published nor shared beyond the project team. The data will only be published through aggregated statistics and in anonymized form and thus no conclusions can be drawn about individuals.\\
The anonymized data will be stored on a password-protected encrypted computer and a private, password-protected online repository. All anonymized data will be stored for a maximum of two years after the end of the study.\\
\newline

\#\#\#\#\# What are my rights to my personal data?\\
Before the irrevocable anonymization of the collected data, you can request information about the personal data collected from you at any time and without giving reasons. You can also request that it be rectified, handed over to you, barred for processing or erased. To do so, please contact the person indicated above. Please write down and remember your participant ID indicated at the top of this form as it will be required to identify your anonymous data.

The Federal Data Protection Act (FADP), and if applicable the European General Data Protection Regulation (GDPR)) are complied with when collecting and processing personal data.\\
\newline

\#\#\#\#\# How am I insured?\\
You are insured for adverse health effects that are directly caused by the study and can be demonstrated to be attributable to fault on the part of the project team or <REDACTED>. You are responsible for insuring yourself against any other adverse health effects such as might occur, for instance, in connection with the trip to or from the place where the study is conducted.\\
\newline

\#\#\#\#\# Who funds this study? Who reviewed this study?\\ 
The study is funded by <REDACTED>.\\
This study was examined by the <REDACTED>.\\
The secretariat of the <REDACTED> is available to help you with complaints in connection with your participation. Contact: <REDACTED>\\

\#\# Consent Form:\\
I, the participant, confirm by clicking the following button that:\\ 
- I have read and understood the study information. My questions have been answered completely and to my satisfaction.\\ 
- I comply with the inclusion criteria for participation described above. I am aware of the requirements and restrictions to be observed during the study.\\
- I have had enough time to decide about my participation.\\ 
- I participate in this study voluntarily and consent that my personal data be used as described above.\\
- I understand that I can stop participating at any moment.\\
- I have written down my participant ID and understand that it is the only way that the authors can identify my data for any request I can have, as my data is anonymized.\\
\newline
\#\#\#\#\# Participant ID: <ID>\\  

\#\#\#\#\# \textbf{By clicking the button, I confirm that I have read the above information and agree to it.}\\
  \end{framed}
  \caption{Instructions and consent form signed by the field test participants}
  \label{instructions}
\end{figure*}

\subsection{User interface}  %

\begin{figure*}[t]  %
    \centering
    \includegraphics[width=0.8\textwidth]{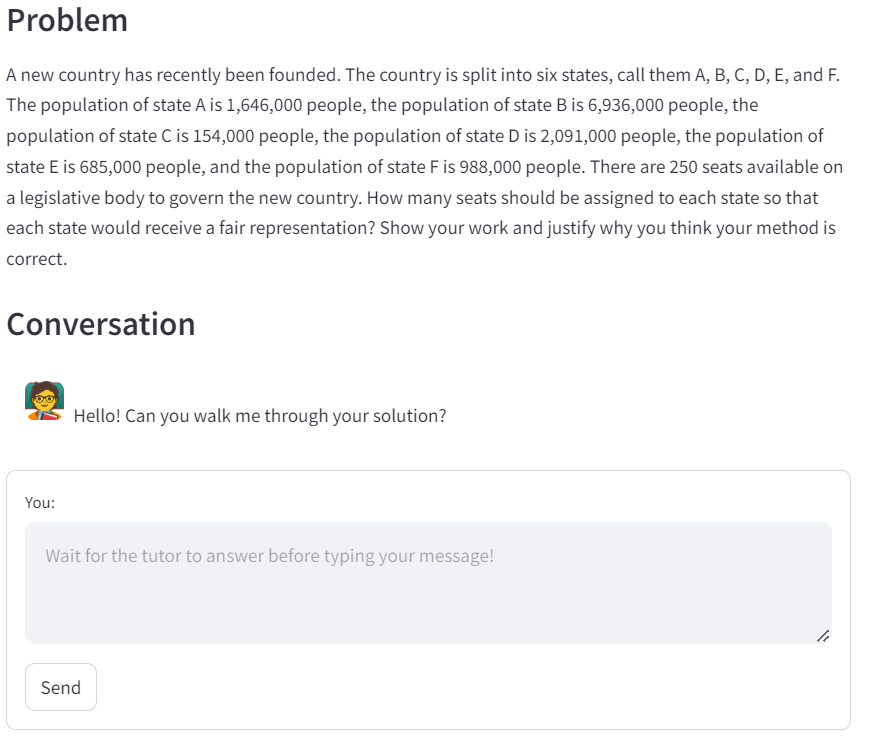}
    \caption{User interface of our math tutor prototype used by students in the field study.}
    \label{fig:user_interface}
\end{figure*}

Figure \ref{fig:user_interface} displays the chat interface of our prototype tutor.

\clearpage

\section{\SECappendix{}}\label{sec:appendix}
We release our code under the MIT License in the supplementary material.
\subsection{\SECappendixtaxonomy{}\label{sec:appendixtaxonomy}}

We introduce a Learning-Sciences--based taxonomy of tutoring intents denoted $\mathcal{I}$ in Section \ref{sec:tutoringstrategymodeling}. An \textit{intent} is defined as the goal of a dialog utterance, such as correcting a mistake. The taxonomy groups all possible intents a tutor implemented with \name\ can have during the tutoring session. We establish this taxonomy using the theories of Scaffolding~\cite{structuring_problematizing_scaffolding}, Affective Cognition~\cite{motivation} and PF~\cite{when_pf_fails,Designing}.

We propose the following four categories of intents:

\begin{enumerate}
    \item Scaffolding. The goal is to restrict the solution space, for example by decomposing the problem in steps, focusing the student's attention on what is important, keeping track of what they did, or offloading unimportant parts of the task such as computations.
    \item Problematizing. This category of intents aims at eliciting a deeper understanding through active engagement with the problem. In contrast to scaffolding, problematizing interventions expand the solution space by asking the student to generalize their answer, consider its limits, or understand its broader implications.
    \item Affective. These intents aim at supporting the student in the often frustrating process of learning, and at drawing their interest.
    \item Generic. The intents in this category serve broader conversational purposes, like greetings.
\end{enumerate}
We subdivide and organize them in the taxonomy in Table \ref{tab:taxonomy}.

\begin{table*}%
  \centering
  \begin{tabularx}{\textwidth}{|>{\raggedright\arraybackslash}p{3.25cm}|X|>{\raggedright\arraybackslash}p{6.5cm}|}
    \hline
    \textbf{Category} & \textbf{Intent $I$} & \textbf{Example} \\
    \hline
    \multirow{6}{3.5cm}{\textbf{Structuring} \cite{structuring_problematizing_scaffolding}}
    & Guide Self-correction & Take a second look at your formula. \\
    \cline{2-3}
    & Correct the Mistake & Are you sure AB is the hypotenuse? \\
    \cline{2-3}
    & Ask for the Next Step & How would you proceed from now? \\
    \cline{2-3}
    & Give a Hint & Try to use the Pythagorean theorem. \\
    \cline{2-3}
    & State & The Pythagorean theorem states that \ldots. \\
    \cline{2-3}
    & Perform a Computation & The result of this computation is \ldots. \\
    \hline
    \multirow{4}{3.25cm}{\textbf{Problematizing} \cite{structuring_problematizing_scaffolding}}
    & Identify Limits & Can you think of a case where it fails? \\
    \cline{2-3}
    & Ask for Intuition & In your own words, what could be \ldots. \\
    \cline{2-3}
    & Ask for Articulation & Can you formally write this idea? \\
    \cline{2-3}
    & Guide Self-Reflection \cite{reflective} & What is the main takeaway of today? \\
    \hline
    \multirow{4}{3.25cm}{\textbf{Affective} \cite{motivation,when_pf_fails}}
    & Maintain Sense of Challenge & I dare you to compute \ldots. \\
    \cline{2-3}
    & Bolster Self-Confidence & Seems like you outsmarted me! \\
    \cline{2-3}
    & Promote Sense of Control & How do you want to call this triangle? \\
    \cline{2-3}
    & Evoke Curiosity & These equations are also used in \ldots. \\
    \hline
    \multirow{2}{3.25cm}{\textbf{Generic}~\cite{mathdial}}
    & Say Goodbye & Congratulations! See you next time. \\
    \cline{2-3}
    & Other & Yes, you are allowed to use a calculator. \\
    \hline
  \end{tabularx}
  \caption{Learning Sciences--based taxonomy of tutor intents.}
  \label{tab:taxonomy}
\end{table*}

\subsection{\SECappendiximplementation{}\label{sec:appendiximplementation}}

\subsubsection{\SECappendixassessor}\label{sec:appendixassessor}
The set $\mathcal{S}$ (as per the notation used in Section \ref{sec:studentstatespace}) of labels (also called features) we choose from during the Student State Tracing procedure have been discussed with learning scientists and the prompt structure follows one used by the CLASS framework~\cite{SPOCK}. We implement the assessment as a multi-label natural language classification task: the last student's utterance is mapped to a predefined set of features representing the student's intents, errors' type, and emotional state. The full list of features can be read from the prompt in Figure~\ref{fig:Assessor_prompt} (letters a--m in the prompt). We call GPT-4o version `gpt-4o-2024-08-06' on this prompt, with a \texttt{temperature} of 0 and a \texttt{top\_p} parameter of $0.1$.

\begin{figure}[h!]
    \begin{lstlisting}[numbers=right,basicstyle=\ttfamily\tiny, numberstyle=\tiny\ttfamily, breaklines=true]
A student and their tutor are working on a math problem:
*Problem Statement*:
##
{pb}
##

The *Provided Solution* of this problem is:
##
{sol}
##

The tutor's utterances are preceded by "Tutor:" and the student's utterances are preceded by "Student:".

Analyze the last student's utterance.
select all the feedbacks that apply from "a,b,c,d,e,f,g,h,i,j,k":

a) The student is using or suggesting a wrong method or taking a wrong path to solve the problem
b) The student made an error in the algebraic manipulation
c) The student made a numerical error
d) The student provided an intuitive or incomplete solution
e) The student's answer is not clear or ambiguous
f) The student correctly answered the tutor's previous question
g) The student is explicitly asking about how to solve the problem
h) The student is explicitly asking the tutor to state a specific theorem
i) The student is explicitly asking the tutor to do a numerical calculation
j) The student and tutor arrived at a complete solution for the entirety of the initial *Problem Statement*
k) The student and tutor arrived at a complete solution for the entirety of the initial *Problem Statement* equivalent to the method provided in the *Provided Solution*

Moreover select if relevant some emotional states from "l,m":
l) The student shows a strong lack of motivation
m) The student shows a strong lack of self-confidence

Proceed step by step. First briefly justify your selection, then provide a string containing the selected letters.
Answer in the following json format:
##
{{
    "justification": "..",
    "selection": ".."

}}
##
Analyze the last student's utterance.
{
    \end{lstlisting}
    \caption{Student State Tracing Prompt to select the student state features, from those enumerated with lowercase letters (`\texttt{a}', ..., `\texttt{m}').}
    \label{fig:Assessor_prompt}
\end{figure}

We require the model to generate a justification for its assessment before writing the final selection. This justification serves a double purpose. First, it enhances the system's explainability. We used these justifications to understand why the LLM selected some erroneous labels and used them to incrementally improve the system, including the formulation of the prompt. Second, this technique usually enhances the accuracy of classification task~\cite{justification_Assessor}. For this purpose, it is important to require the justification to be written \textit{before} the final label selection, as we use an autoregressive LLM, which can only consider context monodirectionally from the left.
In the prompt, we use `\#\#' as a delimiter between the instructions and grounding information (such as the problem and its solution), as suggested in~\cite{Official_guide}.
In lines 2 and 7 we surround ``Problem Statement'' and ``Provided Solution'' by `*' to facilitate their reference later in the prompt~\cite{prompt_programming}. The core selection prompt from lines 15 to 31 is inspired by the assessment used in the CLASS framework~\cite{SPOCK}. The exact formulation for each label has been optimized by trial and error over development iterations of the prototype. Lines 33 and 42 summarize the goal and ask the LLM to work step-by-step, a prompting technique drastically improving performance on reasoning benchmarks~\cite{zero_shot}.
The last line, containing the first character of the expected answer (`\{'), acts as a syntactical constraint to guide the LLM to follow the JSON format~\cite{prompt_programming}.

The LLM also keeps in history the past utterances and their corresponding assessments.

\begin{table*}[t]
\scriptsize
  \centering
  \begin{tabularx}{\textwidth}{|p{3.25cm}|X|p{7cm}|}
    \hline
    \textbf{Category} & \textbf{Intent} & \textbf{Text of the prompt addition} \\
    \hline
     \multirow{3}{2.5cm}{\textbf{Structuring}}
    & Guide Self-correction & Make the student identify by themself the error in their answer.\\
    \cline{2-3}
    & Correct the Mistake & Correct the student's mistake by giving them a hint.\\
    \cline{2-3}
    & Ask for Intuition & Encourage and make the student find on their own what is the next step to solve the problem, for example by asking a question.\\
    \cline{2-3}
    & Give a Hint & Give a hint to the student to help them find the next step. Do *not* provide the answer.\\
    \cline{2-3}
    & State & State the theorem or definition the student is asking about.\\
    \cline{2-3}
    & Perform a Computation & Correct and perform the numerical computation for the student.\\
    \hline
    \multirow{2}{2.5cm}{\textbf{Problematizing}} 
    & Identify Limits & Make the student identify the limits of their reasoning or answer by asking them questions. \\
    \cline{2-3}
    & Ask for Intuition & Ask the student start by providing a guess or explain their intuition of the problem.\\
    \cline{2-3}
    & Ask for Articulation & Ask the student to write their intuition mathematically or detail their answer.\\
    \cline{2-3}
    & Guide Self-Reflection & Step back and reflect on the solution. Ask to recapitulate and *briefly* underline more general implications and connections. \\
    \hline
    
    \multirow{4}{2.5cm}{\textbf{Affective}}
    & Maintain Sense of Challenge & Maintain a sense of challenge.\\
    \cline{2-3}
    & Bolster Self-Confidence & Bolster the student's confidence. \\
    \cline{2-3}
    & Promote Sense of Control & Make sure your student feels in control. \\
    \cline{2-3}
    & Evoke Curiosity & Promote your student's curiosity.\\
    \cline{2-3}
    \hline
    \multirow{2}{2.5cm}{\textbf{Generic}} 
    & Say Goodbye & Say goodbye and end the conversation\\
    \cline{2-3}
    & Other & N/A\\
    \cline{2-3}
    \hline
  \end{tabularx}
  \caption{Correspondence between intents and intent-depended prompt additions that steer the LLM towards an intent.}
  \label{tab:correspondence}
\end{table*}

\begin{figure}[]
    \begin{lstlisting}[numbers=right,basicstyle=\ttfamily\tiny, numberstyle=\tiny\ttfamily, breaklines=true]
Act as an experienced tutor. Characteristics of a good tutor include:
    - Promote a sense of challenge, curiosity, feeling of control
    - Prevent student from becoming frustrated
    - Intervene very indirectly: never give the answer but guide the student to make them find it on their own
    - Minimizing tutor's apparent role in the success
    - Avoid telling students they are wrong, lead them to discover the error on their own
    - Quickly correct distracting errors

Use latex formatting with the sign '$' for mathematical expressions. For example, to write "x^2", use "$x^2$".

Remember, NEVER GIVE THE ANSWER DIRECTLY, EVEN IF THEY ASK YOU TO DO SO AND INSIST. Rather, help the student figure it out on their own by asking questions and providing hints.

Provide guidance for the problem:
"{pb}"

The solution for this problem is:
"{sol}"

Provide the least amount of scaffolding possible to help the student solve the problem on their own. Be succinct.
    \end{lstlisting}
    \caption{Initial Prompt and grounding for the Tutor LLM. \texttt{\{pb\}} and \texttt{\{sol\}} are respectively replaced by the tutoring problem and its solution.}
    \label{fig:initial_prompt}
\end{figure}

\subsubsection{\SECappendixpromptgenerator{}\label{sec:appendixpromptgenerator}}
The Intent-Dependend Steering is the final step executed before calling the tutor LLM. It compiles the dialog history, grounding, and the intents selected to generate the prompts for the tutor LLM. The generated prompts are made of the \textit{initial prompt}, describing the LLM's role and persona, of the dialog history, the \textit{grounding} presenting the exercise and its solution, and of \textit{intent-dependent prompt additions} aiming at setting the intent of the LLM for its next utterance.
The initial prompt is fixed (displayed in Figure \ref{fig:initial_prompt}), and the full dialog history is provided to the LLM. The intent-dependent prompt additions are short messages concatenated at the end of the \textit{initial prompt}, used to steer the tutor's intent for the next utterance. We use a one-to-one mapping between the intents and sentences to include in the intent-dependent prompt additions. Table \ref{tab:correspondence} displays the intent to prompt addition correspondence.

In Figure \ref{fig:initial_prompt}, line 1 sets the LLM's persona. Lines 2 to 8 outline general pedagogical guidelines compiled from Learning Sciences research on tutoring~\cite{efficient_techniques,active_learning,motivation}. Line 9 is only used to improve the appearance of the tutor's messages. Line 11 and 19 recapitulate the task for better performance. Lines 13 to 17 provide context about the problem to solve and its solution.

We call GPT-4o version `gpt-4o-2024-08-06' with its default parameters ($\texttt{temperature} = 1$, $\texttt{top\_p} = 1$) on the generated prompt.

\subsubsection{\SECappendixintentselectorprompt\label{sec:appendixintentselectorprompt}}
Figure~\ref{fig:LLM_Intent_Selector_Prompt} displays the prompt of the LLM used for the Intent Selection in the baseline V4.

\begin{figure}[htbp]
    \begin{lstlisting}[numbers=left,basicstyle=\ttfamily\tiny, numberstyle=\tiny\ttfamily, breaklines=true]
A student and a tutor are working on a math problem. The tutor responds to the student's messages following some intents. Before responding, the tutor first selects the most appropriate intents from a *Taxonomy*.
The tutor selects the intents based on the PF teaching method. The goal is to let the student explore the space of solution by themself to generate as many representation and solution methods as possible, even if they are only partially correct.
Select the best set of intents for the tutor's next message based on the following *Taxonomy*:

*Taxonomy*:
##
    P_LIMITS : Ask the student to identify some limits of their reasoning or answer.
    P_HYPOTHESIS : Ask the student to start by providing a guess, hypothesis or explain their intuition of the problem.
    P_ARTICULATION : Ask the student how they can write their intuition mathematically or detail their answer.
    P_REFLECTION : Reflect on the solution, recapitulate, and underline more general implications.
    S_SELFCORRECTION : Ask a question to hint the student to identify errors in their answer.
    S_CORRECTION : Correct a student's mistake
    S_STRATEGY : Make the student find and perform the next step to solve the problem without giving a hint.
    S_State : State a theorem or definition asked by the student.
    S_OFFLOAD : Perform a computation for the student
    S_HINT : Give a hint to the student
    G_GREETINGS : Say goodbye to the student
##

The tutor's previous intents for their last message to the student were: {previous_intent}.
The student answered the last tutor's message. An assessment of their answer has been conducted using the following codes:

assessment codes:
##
a) The student is using or suggesting a wrong method or taking a wrong path to solve the problem
b) The student made an error in the algebraic manipulation
c) The student made a numerical error
d) The student provided an intuitive or incomplete solution
e) The student's answer is not clear or ambiguous
f) The student correctly answered the tutor's previous question
g) The student is explicitly asking about how to solve the problem
h) The student is explicitly asking the tutor to state a specific theorem
i) The student is explicitly asking the tutor to do a numerical calculation
j) The student and tutor arrived at a complete solution for the entirety of the initial *Problem Statement*
k) The student and tutor arrived at a complete solution for the entirety of the initial *Problem Statement* equivalent to the method provided in the *Provided Solution*
l) The student shows a strong lack of motivation
m) The student shows a strong lack of self-confidence
##

The assessment for the last student's message is: {assessment_codes}.

Based on the assessment and the last tutor's intents, select the most appropriate set of intents for the next tutor's message.
Proceed step by step. First, briefly justify your selection, then provide a list containing the selected intents from the *Taxonomy*.

Answer in the following json format:
##
{
    "justification": "...",
    "intents": []

}
##
{
    \end{lstlisting}
    \caption{Prompt for the LLM Intent Selection in V4.}
    \label{fig:LLM_Intent_Selector_Prompt}
\end{figure}

\subsection{Learning Sciences Rationales of our PF Modelling\label{sec:appendixlearningsciencesbases}} The Intent Selection graph presented in Figure~\ref{fig:transition_graph} has been created from expert knowledge in PF. We detail in this section some of the principles we used in its design.

In PF, the goal is to make the student explore the solution space of an open-ended problem as much as possible, even if they fail to produce a complete solution. This idea gave its name to Productive \textit{Failure}: there is, in this case, a hidden efficacy to the student's failure~\cite{productive_failure}.
To this aim, in Figure \ref{fig:transition_graph} any mistake (coded by features `\texttt{a}', `\texttt{b}' or `\texttt{c}') leads to a `\texttt{Guide Self Correction}' intent. The aspiration is to make the student spot and correct their mistake by themself. If the error persists, we use the intent `\texttt{Correct the Mistake}' that gives a strong hint to the student about their mistake, but without directly correcting it. If the mistake is computational only, we use the intent `\texttt{Perform a Computation}'. This allows the student to focus on the reasoning rather than spending time on computations. For the same reason, if the student asks for a basic theorem or definition they forgot (feature `\texttt{h}'), we use the intent `\texttt{State}' to answer their question.

A good tutor should intervene very indirectly~\cite{efficient_techniques}: a good principle to follow as a tutor is \textit{to give as little scaffolding as one can get away with}. The intent used whenever a student manages to correct their mistake (arrow labeled `\texttt{f}' leaving the `\texttt{Guide Self-Correction}' node in Figure \ref{fig:transition_graph}) or when the student suggests an intuitive or incomplete solution (state feature `\texttt{d}') is the `\texttt{Ask for the Next Step}' intent. This encourages the student to pursue their line of thought without heavier scaffolding.

One of the challenges of PF is to create a classroom culture that allows for failure and promotes psychological safety. Indeed, as students deal with open-ended problems before being given the tools to solve them, the experience can be stressful. In traditional classroom settings, the focus is directed towards getting the right solution and students are often evaluated on this with a grade. One of the most crucial design features of a PF problem session is to shift this goal from solving the problem to generating multiple RSMs regardless of their correctness~\cite{Designing}. As explained in \citet{Designing}, ``students are used to seeking assistance from their teachers so much so that they do so even before sufficiently trying to solve problems themselves. At the same time, teachers are just as used to providing assistance when it is asked for so much so that often opportunities for students to generate and explore RSMs are missed''. For this reason, in our modeling, if the student explicitly asks for help the intent `\texttt{Ask for Intuition}' is triggered. This intent encourages the student to start with a guess or an intuition with no fear of being incorrect. Once the student is successful in providing an intuitive solution, or if they confidently propose an incomplete one (feature `\texttt{e}'), then the `\texttt{Ask for Articulation}' intent is used to encourage an RSM creation by formulating their idea in concrete math. To mitigate stress, that can hinder the student's learning~\cite{motivation}, we trigger a `\texttt{Give a Hint}' intent to ease the student's progress if they show signs of lack of motivation or self-confidence.

A good PF problem allows for multiple solutions (and often a preferred one, called canonical solution). Then, if the student manages to propose a non-canonical solution to the problem (feature `\texttt{j}'), they should be encouraged to continue exploring.
The `\texttt{Identify Limits}' intent is triggered to ask the student to identify the limitations of their solution, encouraging them to develop a new one.

If the student reaches the canonical solution (feature `\texttt{k}'), we use the intent `\texttt{Guide Self-Reflection}'. It has been shown that it is beneficial for learning to engage in self-reflection (i.e. looking back critically at the work done) after a problem session~\cite{reflective}. Following this turn, the tutor is made ready to end the conversation through the `\texttt{Say Goodbye}' intent.

\end{document}